\documentclass[%
reprint,
superscriptaddress,
 amsmath,amssymb,
 aps,
prl,
longbibliography,
]{revtex4-1}
\usepackage{textcomp}
\usepackage{graphicx}
\usepackage{dcolumn}
\usepackage{bm}
\usepackage{color}
\usepackage{lineno}

\begin{document}

\preprint{APS/123-QED}

\title{Microwave one-way transparency by large synthetic motion of magnetochiral polaritons in metamolecules}

\author{Kentaro Mita}
\affiliation{Department of Physics, Graduate School of Science, Tohoku University, Sendai 980-8578, Japan}

\author{Toshiyuki Kodama}
\affiliation{Institute for Excellence in Higher Education, Tohoku University, Sendai 980-8576, Japan}
\affiliation{Organization for Advanced Studies, Tohoku University, Sendai 980-8576, Japan}

\author{Toshihiro Nakanishi}
\affiliation{Department of Electronic Science and Engineering, Kyoto University, Kyoto 615-8510, Japan}

\author{Tetsuya Ueda}
\affiliation{Department of Electrical Engineering and Electronics, Kyoto Institute of Technology, Kyoto 606-8585, Japan}

\author{Kei Sawada}
\affiliation{RIKEN SPring-8 Center, Sayo 679-5148, Japan}

\author{Takahiro Chiba}
\affiliation{Department of Information Science and Technology, Graduate School of Science and Engineering, Yamagata University, Yonezawa 992-8510, Japan}
\affiliation{Department of Applied Physics, Graduate School of Engineering, Tohoku University, Sendai 980-8579, Japan}

\author{Satoshi Tomita}
\email[Email address:]{tomita@tohoku.ac.jp}
\affiliation{Department of Physics, Graduate School of Science, Tohoku University, Sendai 980-8578, Japan}
\affiliation{Institute for Excellence in Higher Education, Tohoku University, Sendai 980-8576, Japan}

\date{\today}

\begin{abstract}
We observe microwave nonreciprocal one-way transparency 
via ultrastrongly-coupled magnetochiral polaritons (MChPs) 
in a metamolecule at room temperature. 
The experimental results using MCh metamolecules 
with simultaneous breaking of time-reversal and space-inversion symmetries 
are reproduced by numerical simulations.
Based on effective polarizability tensor analyses, 
we verify massive synthetic motion of MChPs 
as an origin of the one-way transparency.
This study paves a way to hybrid quantum systems 
and synthetic gauge fields using metamaterials.
\end{abstract}

\maketitle

\textit{Introduction} -- 
Condensed matter physics is science of quasiparticles.
Indeed, 
a lot of complicated phenomena in matters can be understood 
as simple movements of small numbers of quasiparticles with elementary excitations.
Generating and identifying 
complex quasiparticles  
is thus a classical issue \cite{Mills1974}, 
but still at the forefront of condensed matter physics 
and materials science \cite{Kockum2019,Diaz2019,Qin2024}.
In this regard,
it comes as no surprise that 
magnon polaritons (MPs) \cite{Soykal2010}, 
in which quantized spin waves  (magnons) in magnetic materials
are coupled to photons, 
stimulate a flurry of current interest \cite{Huebl2013,Tabuchi2014,Li2020,Harder2021}
and hold great promise 
for realizing novel spintronic hybrid quantum systems \cite{Lachance-Quirion2019}.
Common MPs generated in magnetic materials
in metallic or superconducting cavities for microwaves 
are reciprocal \cite{Golovchanskiy2021}.
However, 
nonreciprocity is indispensable 
in sensitive signal detection and processing, 
particularly in the quantum regime \cite{Walls_book}.

Directionally nonreciprocal MPs with polarization plane rotation, 
which is similar to magneto-optical (MO) effects 
commonly observed in magnetic-material-based devices, 
are obtained using natural breaking of time-reversal symmetry in magnetic materials
in the cavity magnonics systems \cite{Wang2019,Zhang2020}. 
Furthermore,
man-made magnetochiral (MCh) metamaterials 
with engineered breaking of space-inversion symmetry 
in addition to broken time-reversal symmetry \cite{Tomita2014,Tomita2017,Tomita2018,Kurosawa2022,mita2025} 
hint at the likelihood of 
another class of directional nonreciprocity 
of ultrastrongly-coupled MPs.
This class of nonreciprocity without polarization rotation
is referred to as synthetic moving effects \cite{Huidobro2019,asadchy2020}, 
which include
optical MCh effects \cite{Barron1984,Rikken1997}, 
optical magnetoelectric (ME) effects \cite{Kezsmarki2014,Kezsmarki2015,Toyoda2015}, 
and directional birefringence 
in natural chiral molecules under magnetic fields 
and multiferroic materials \cite{Kuzmenko2015,nagaosa2024}.
An enhancement in the synthetic moving effects, 
which results in microwave one-way transparency, 
is of great importance 
in application to 
hybrid quantum systems for quantum computing \cite{Lachance-Quirion2019} 
and to synthetic Lorentz force acting on microwaves \cite{Sawada05}.
Nevertheless, 
the mechanism of 
the enhancement in the synthetic motion
is still an open question \cite{Kurosawa2022,mita2025}. 

In this Letter, 
we have achieved microwave nonreciprocal one-way transparency 
using a single MCh metamolecule in a waveguide
under moderate magnetic fields 
at room temperature. 
The MCh metamolecule consists of 
a polycrystalline yttrium-iron garnet (YIG) cylinder 
as a magnetic meta-atom 
inserted in a right-handed helix made of copper (Cu)  
as a chiral meta-atom. 
Microwave transmission spectra 
of the MCh metamolecule
under direct current (DC) magnetic fields
demonstrate ultrastrongly-coupled MPs,
to which we refer as MCh polaritons (MChPs), 
and microwave one-way transparency. 
These experimental results 
are reproduced via numerical simulations.
More strikingly, 
based on effective polarizability tensor analyses 
\cite{Mirmoosa2014,Alaee2015,Yazdi2016,Kodama2024}, 
we reveal the enhancement mechanism of the synthetic motion, 
in which the combination of chiral-type bianisotropy and large MO effects 
causes massive synthetic motion of MChPs 
for microwave one-way transparency.
This mechanism of enhancement in the synthetic motion 
is applicable to natural chiral molecules \cite{Rikken1997} 
and multiferroic materials \cite{Kezsmarki2014,Kezsmarki2015,Toyoda2015}.

\begin{figure}[tb!]
\includegraphics[width=7.5truecm,clip]{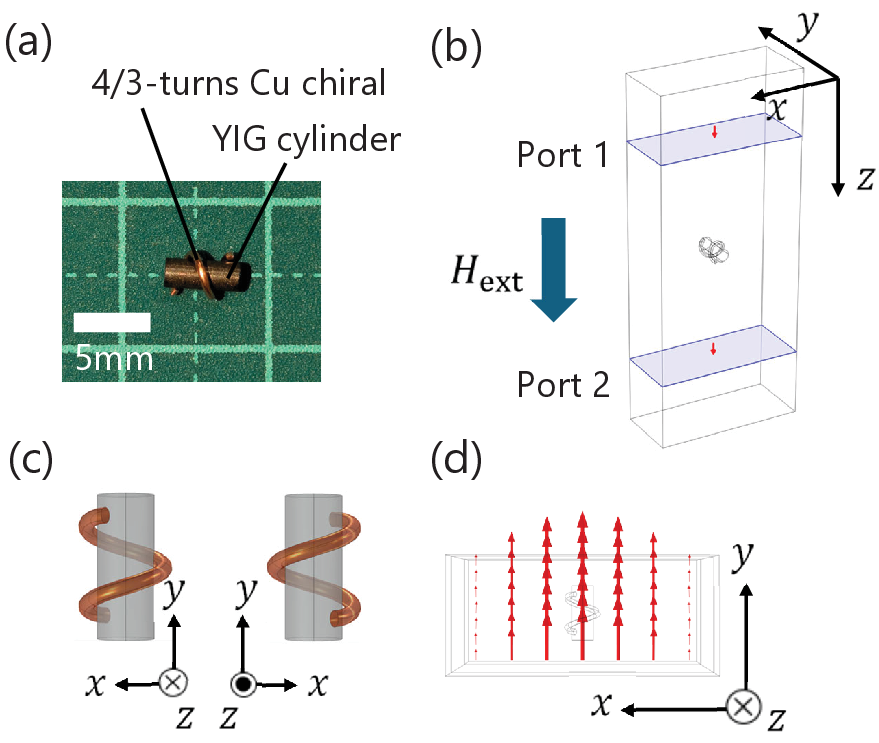}
\caption{
(a) A photo of the MCh metamolecule 
consisting of YIG magnetic meta-atom and Cu chiral meta-atom. 
A white bar corresponds to 5 mm.
(b) Microwave measurement setup. 
The metamolecule is oriented along the $y$-axis in the WR-90 waveguide. 
A DC magnetic field $\mu_{0} H_{\rm ext}$
up to 500 mT 
is applied in the +$z$ direction 
using an electromagnet. 
(c) $x$-$y$ plane views of the metamolecule. 
The coil's endpoints are arranged 
to have 180-degree rotational symmetry 
with respect to the $x$-axis.
(d) A cross-sectional view in the $x$-$y$ plane of the WR-90 waveguide. 
Red arrows correspond to the AC electric fields of the TE$_{10}$ mode.
}
\label{fig:sample}
\end{figure}

\textit{Experimental setup} --
Figure \ref{fig:sample}(a) presents 
a photo of the MCh metamolecule. 
The Cu wire of 0.55 mm diameter 
is wounded 4/3 times to form the right-handed helix. 
The length is one-third 
compared to that in the previous studies \cite{Tomita2014,Tomita2017,Tomita2018,mita2025}
to measure the microwave transmission 
of the metamolecule 
at different orientation in the waveguide, 
while the pitch (2.6 mm) and outer diameter (3.10 mm) of the helix 
are the same to show sharp chiral resonance at approximately 9 GHz.
The YIG cylinder 
is 5 and 2 mm in length and diameter, respectively. 

As illustrated in Fig. \ref{fig:sample}(b),
a single MCh metamolecule 
is set into a WR-90 waveguide 
that supports the TE$_{10}$ mode 
with square flange adapters (Pasternack PE9804). 
The metamolecule is oriented along the $y$-axis in Fig. \ref{fig:sample}(b). 
The DC magnetic field $\mu_{0} H_{\rm ext}$ 
up to 500 mT 
is applied in the +$z$-direction
using an electromagnet. 
Figure \ref{fig:sample}(c) 
shows cross-sectional views in the $x$-$y$ plane 
of the metamolecule. 
The coil's endpoints are arranged 
to have 180-degree rotational symmetry 
with respect to the $x$-axis.

The waveguide is connected to 
a vector network analyzer (VNA) 
(Rohde \& Schwarz ZVA67) 
with a microwave input power of 0 dBm (1 mW). 
Figure \ref{fig:sample}(d) 
presents a cross-sectional view in the $x$-$y$ plane
of the WR-90 waveguide. 
Red arrows correspond to 
AC electric fields of the TE$_{10}$ mode microwave photon.
Along to the $y$-axis, 
there are the AC electric fields 
but no AC magnetic fields.  
The AC electric fields in the $y$-direction
and external $\mu_{0} H_{\rm ext}$ in the $z$-direction 
give rise to MChPs with large synthetic motion 
in the $y$-axis oriented metamolecule.

The complex scattering parameters, so-called $S$ parameters, 
are measured using VNA. 
The $S_{21}$ represents a complex transmission coefficient from port 1 to 2, 
while $S_{12}$ indicates that from port 2 to 1. 
$|S_{21}|^2$ corresponds 
to transmittance $T^{+}$ of microwaves propagating in the $+z$-direction, 
whereas $|S_{12}|^2$ corresponds 
to transmittance $T^{-}$ of microwaves propagating in the $-z$-direction. 
All measurements are carried out at room temperature.

\begin{figure}[tb!]
\includegraphics[width=8.5truecm,clip]{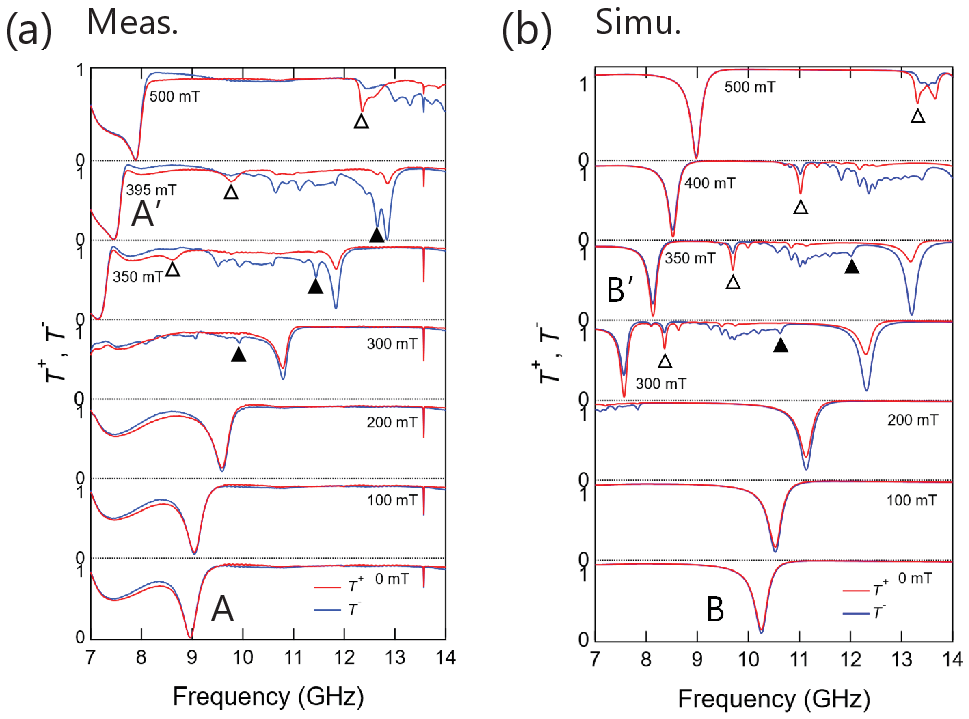}
\caption{
(a) Measured $T^{+}$ (red) and $T^{-}$ (blue) spectra 
at various external DC magnetic fields, $\mu_{0} H_{\rm ext}$
of 0 - 500 mT.
Black and white triangles indicate dips 
caused by Mie-coupled MChPs.
(b) Numerically calculated $T^{+}$ (red) and $T^{-}$ (blue) spectra 
at various external DC magnetic fields of 0 - 500 mT. 
The phenomenological Gilbert damping parameter $\alpha$ is set to 0.004 
in the calculation.
}
\label{fig:s21s12}
\end{figure}

\textit{Measurement and simulation results} --
Figure \ref{fig:s21s12}(a) presents 
microwave transmittance  
$T^{+}$ (red) and $T^{-}$ (blue) spectra 
measured for the MCh metamolecule
at various $\mu_{0} H_{\rm ext}$ of 0 - 500 mT. 
At $\mu_{0} H_{\rm ext} = 0$ mT, 
the $T^{+}$ and $T^{-}$ spectra 
are identical 
and show a dip at 9.0 GHz labelled as A. 
Because the AC electric fields 
are along to the $y$-axis 
as shown in Fig. \ref{fig:sample}(d), 
the dip A is assigned to  
the electric dipole resonance in the Cu chiral meta-atom 
by microwave photons, 
referred to as chiral resonance.
Numerical calculation using COMSOL Multiphysics
indicates that 
the chiral resonance
observed at around 9 GHz in this study 
corresponds to 
the fundamental $n = 1$ resonance mode 
in the Cu chiral structure with 4/3 turns. 
The Supplemental Material \cite{SM} 
details the numerical simulation method.

As $\mu_{0} H_{\rm ext}$ increases to 100, 200, and 300 mT, 
the dip frequency shifts upward.
The blue shift of the chiral resonance 
corresponds to the initial stage of anti-crossing of MChPs 
generated by applied $\mu_{0} H_{\rm ext}$.
Moreover, 
the dip of the $T^{+}$ spectrum (red) 
becomes shallower than that of the $T^{-}$ spectrum (blue).
The difference between $T^{+}$ and $T^{-}$ 
corresponds to nonreciprocity 
due to the synthetic motion of MChPs 
in the MCh metamolecule 
with simultaneously broken space-inversion and time-reversal symmetries. 

At $\mu_{0} H_{\rm ext} =$ 300 mT, 
small dips appear below the chiral resonance A 
as typically indicated by a black triangle on the $T^{-}$ spectrum.
With a further increase in $\mu_{0} H_{\rm ext}$ to 350 mT, 
the small dips shift to higher frequencies. 
In addition, 
another small dip appears at 8.6 GHz  
as indicated by a white triangle on the $T^{+}$ spectrum.
These small dips shift to higher frequencies continuously 
as $\mu_{0} H_{\rm ext}$ increases, 
demonstrating that 
they are relevant to ferromagnetic resonance (FMR),
i.e., magnons, in the YIG magnetic meta-atom.
Numerical calculation using COMSOL 
indicates that 
these FMR-featured dips 
are assigned to other polariton modes, 
in which Mie-resonance \cite{BH} 
in YIG meta-atom on FMR 
are coupled to the chiral resonance. 
We thus refer to 
the polariton modes 
indicated by black and white triangles in Fig. \ref{fig:s21s12}(a) 
as Mie-coupled MChPs (MC-MChPs).
In more detail, see Fig. S1 and Movie S1 in the Supplemental Material \cite{SM}.

At $\mu_{0} H_{\rm ext} =$ 400 mT, 
a large dip labelled as A' 
appears at a lower frequency around 7.5 GHz. 
The large dip A' is traced back to 
MChPs, 
which is shifted to a lower frequency 
owing to the Rabi-like level splitting; 
this is the signature of 
coherent coupling between 
the chiral resonance 
and magnons, 
generating MChPs.
More strikingly, 
the $T^{+}$ spectrum becomes flat 
while the $T^{-}$ spectrum shows 
a dip due to MChPs and fine structures due to MC-MChPs
between 10 and 13 GHz.
This corresponds to 
one-way transparency 
of the $T^{+}$ signal.
With a further increase in $\mu_{0} H_{\rm ext}$ to 500 mT, 
fine structures owing to MC-MChPs indicated by white and black triangles
and a large dip (A) due to MChPs 
keep moving to higher frequencies.
At 12.4 GHz, as indicated by a white triangle, 
the $T^{-}$ spectrum becomes flat 
while the $T^{+}$ shows a dip due to MC-MChPs. 

Figure \ref{fig:s21s12}(b) shows
corresponding simulation results using COMSOL 
with the phenomenological Gilbert damping parameter $\alpha$ of 0.004. 
In Fig. \ref{fig:s21s12}(b), 
red and blue curves correspond respectively 
to calculated $T^{+}$ and $T^{-}$ spectra 
of the $y$-axis-oriented MCh metamolecule in the waveguide 
at various $\mu_{0} H_{\rm ext}$ of 0 - 500 mT.
The numerical simulations 
reproduced qualitatively the experimental results 
in terms of 
the shift of 
the chiral resonance 
labelled as B and B', 
the generation of MChPs with the Rabi-like splitting, 
the generation of MC-MChPs indicated by white and black triangles,
and microwave one-way transparency of MChP 
at 13.2 GHz 
under $\mu_{0} H_{\rm ext} =$ 350 mT.

\begin{figure}[tb!]
\includegraphics[width=8.5truecm,clip]{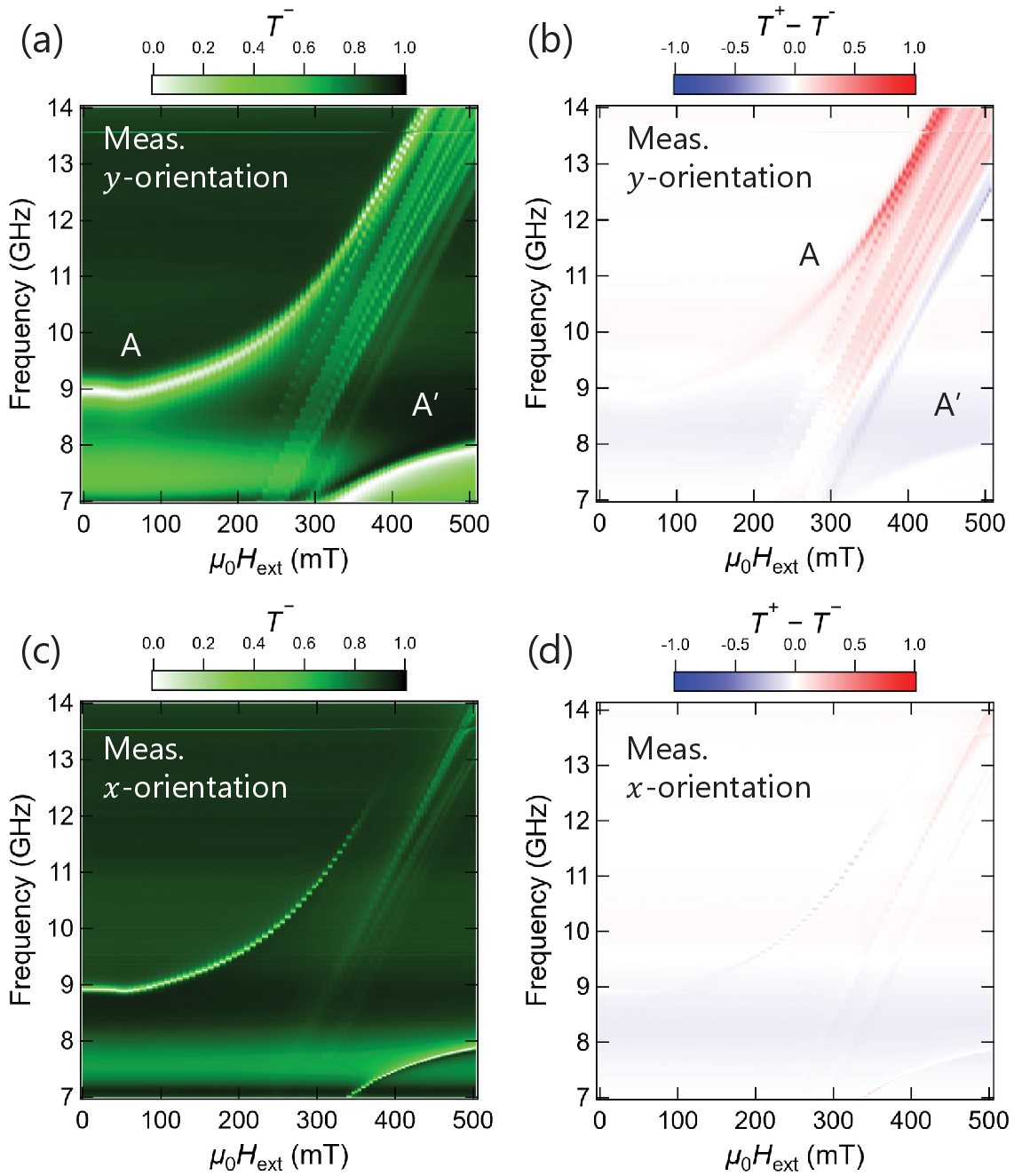}
\caption{2D plots of experimentally observed 
(a) $T^{-}$ and 
(b) $T^{+} - T^{-}$ 
of the $y$-axis-oriented metamolecule, 
and 
(c) $T^{-}$ and 
(d) $T^{+} - T^{-}$ 
of the $x$-axis-oriented metamolecule
as a function of $\mu_{0} H_{\rm ext}$ (horizontal) and frequency (vertical).
}
\label{fig:2dplot}
\end{figure}

\textit{Polariton coupling ratio} -- 
Figure \ref{fig:2dplot}(a) 
presents 
two dimensional (2D) plots of experimentally measured $T^{-}$ 
as a function of $\mu_{0} H_{\rm ext}$ (horizontal) and frequency (vertical).
Dark green color corresponds to $T^{-} =$ 1.0 
while white color corresponds to $T^{-} =$ 0.0.
The coupling ratio of MChPs
are evaluated 
from the Rabi-like level splitting between branches A and A' 
in Fig. \ref{fig:2dplot}(a).
The Rabi-like splitting, $g / \pi$, of MChP
at the intersection 
of the chiral resonance at 9.0 GHz 
and the lower MC-MChP mode (the white triangle in Fig. \ref{fig:s21s12}(a)) 
at $\mu_{0} H_{\rm ext}$ = 364 mT 
is evaluated to be larger than 4.9 GHz.
This brings about a coupling ratio
$g / \omega >$ 0.27, 
which is larger than 0.1, 
indicating MChPs 
in the ultrastrong-coupling regime.  

The coupling ratio 
does not depend on the metamolecule orientation 
in the $x$-$y$ plane.
Metamolecule orientation dependence 
is studied by measuring 
the $x$-axis-oriented MCh metamolecule.
Figure \ref{fig:2dplot}(c) shows 
2D plots of experimentally measured $T^{-}$ 
of the $x$-axis-oriented metamolecule 
as a function of $\mu_{0} H_{\rm ext}$ (horizontal) and frequency (vertical).
The coupling ratio of MPs 
evaluated at 364 mT 
is 0.27, 
which is same as that of MChPs in Fig. \ref{fig:2dplot}(a).
When $\mu_{0} H_{\rm ext}$ is applied in the $z$-direction,
magnetization in the YIG meta-atom 
shows precession in the $x$-$y$ plane. 
Spatial relation 
between the Cu chiral meta-atom 
and the magnetization precession in YIG meta-atom 
is thus identical in spite of metamolecule orientation direction 
in the $x$-$y$ plane.
This leads to the coupling ratio 
that is independent of the metamolecule orientation.
Contrastingly, 
directional nonreciprocity
strongly depends on the metamolecule orientation
as shown in the following.

\textit{Polariton directional nonreciprocity} -- 
The directional nonreciprocity  
is evaluated using 
transmittance difference, $T^{+} - T^{-}$.
When $T^{+} - T^{-} = \pm 1.0$, 
microwaves show perfect one-way transparency.
Figure \ref{fig:2dplot}(b) 
illustrates 
2D plots of $T^{+} - T^{-}$ 
experimentally observed for the $y$-axis-oriented metamolecule 
as a function of $\mu_{0} H_{\rm ext}$ (horizontal) and frequency (vertical). 
In the color scale, 
red corresponds to $T^{+} - T^{-} = 1.0$
while blue corresponds to $T^{+} - T^{-} = -1.0$.
In Figs. \ref{fig:2dplot}(b), 
the upper branch (A) of MChPs 
presents $T^{+} - T^{-} > 0$, 
while the lower branch (A') 
shows  $T^{+} - T^{-} < 0$, 
highlighting  
the nonreciprocity of MChPs and MC-MChPs. 
More interestingly, 
$T^{+} - T^{-} $ is 0.72 
at around 13 GHz by MChP 
at $\mu_{0} H_{\rm ext}$ = 400 mT, 
indicating nearly one-way transparency of microwave 
via MChP.

Figure \ref{fig:2dplot}(b) shows that 
MC-MChP is split into several modes; 
this is consistent 
with split signals 
indicated by black and white triangles in Fig. \ref{fig:s21s12}. 
MC-MChP at a higher frequency 
shows $T^{+} - T^{-} >$ 0
while MC-MChP at a lower frequency 
presents $T^{+} - T^{-} <$ 0; 
the polarity of the nonreciprocity is thus the same 
between MChP and MC-MChP. 
This indicates that 
MC-MChP is relevant to MChP.

Figure \ref{fig:2dplot}(d) presents 
2D plots of $T^{+} - T^{-}$ 
experimentally observed for the $x$-axis-oriented metamolecule.
The nonreciprocity of MChPs 
in the $x$-axis-oriented metamolecule 
shown in Fig. \ref{fig:2dplot}(d)
is much smaller than 
that in the $y$-axis-oriented metamolecule
in Fig. \ref{fig:2dplot}(c).
In this way, 
the nonreciprocity and one-way transparency
strongly depend on the metamolecule orientation.

\textit{Effective polarizability tensor analyses} --
In contrast to the coupling ratio, 
$T^{+} - T^{-}$ strongly depends
on the metamolecule orientation.
In the following, 
using effective polarizability tensor \cite{Mirmoosa2014,Alaee2015,Yazdi2016,Kodama2024}, 
we verify that one-way transparency with large $|T^{+} - T^{-}|$
by the $y$-axis-oriented metamolecule 
is traced back to massive synthetic motion of MChPs. 
Effective polarizability tensor analyses 
\cite{Mirmoosa2014,Alaee2015,Yazdi2016,Kodama2024} 
can be used 
as the metamolecule with outer diameter of 3.10 mm 
is small enough compared to the microwave wavelength, e.g., 30 mm at 10 GHz. 
We numerically calculate microwave $S$ parameters 
of the MCh metamolecules in the free-space with periodic boundary conditions 
using COMSOL.
The calculated $S$ parameters 
are converted to 
reflection and transmission coefficients.
The $T^{+}$ and $T^{-}$ spectra 
for the metamolecule array in the free-space 
(Fig. S2 in Supplemental Material)
is similar to those for the metamolecule in the waveguide (Fig. \ref{fig:s21s12}(b)).  
At $\mu_{0} H_{\rm ext} =$ 400 mT, 
MChP signals are observed 
at 12.2 and 9.3 GHz.
Additionally, 
MC-MChP signal is observed at 11.6 GHz.
The reflection and transmission coefficients 
are converted eventually to
effective polarizability tensors. 
The Supplemental Material \cite{SM} 
details the extraction procedures.

\begin{figure}[tb!]
\includegraphics[width=8.5truecm,clip]{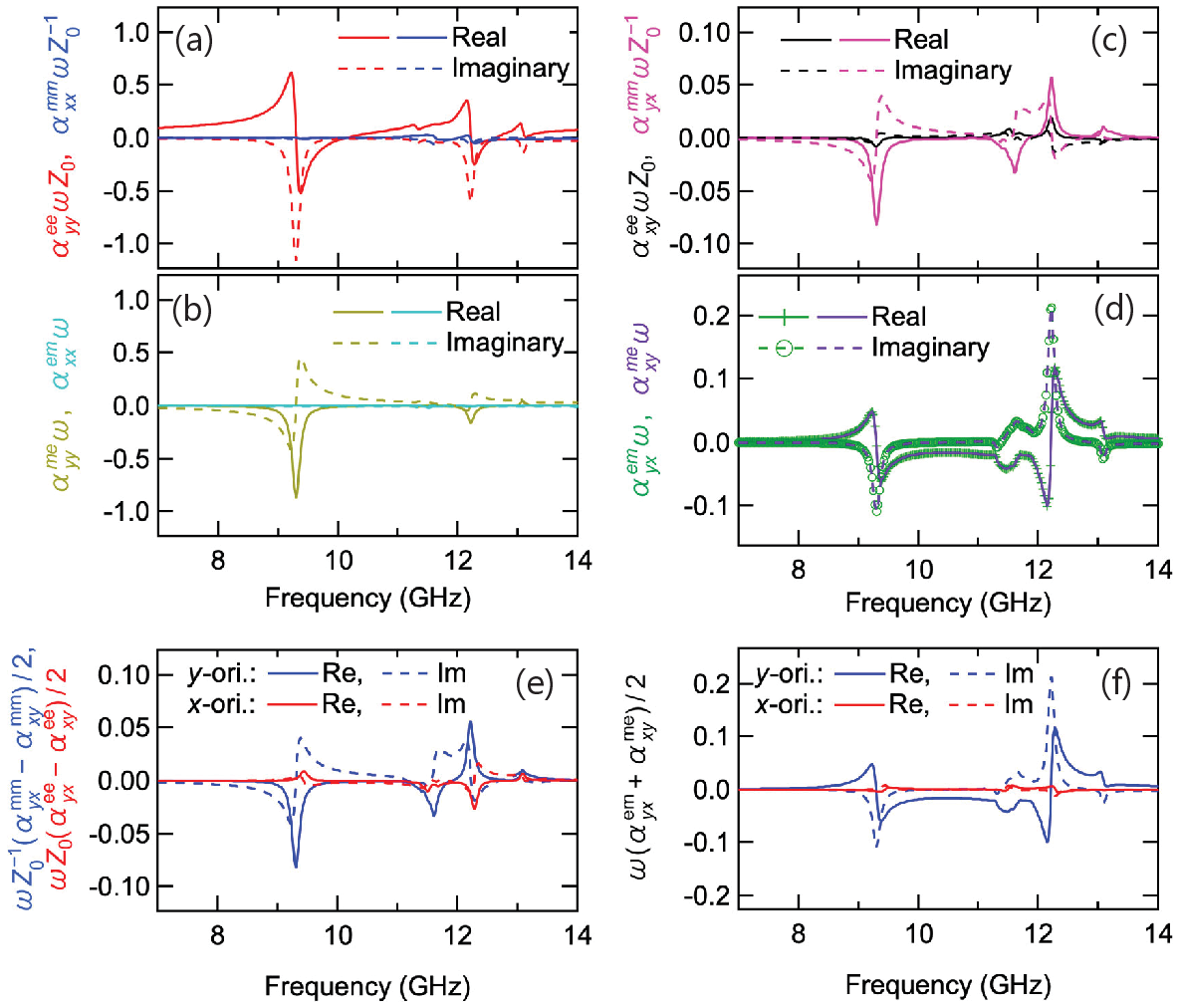}
\caption{
Effective polarizabilities evaluated from numerical simulation 
of the $y$-axis-oriented metamolecule using $y$-polarized waves. 
Frequency versus extracted 
(a) $\alpha_{yy}^{\rm ee} \omega Z_0$ (red) and $\alpha_{xx}^{\rm mm} \omega Z_0^{-1}$ (blue), 
(b) $\alpha_{yy}^{\rm me} \omega$ (gold) and $\alpha_{xx}^{\rm em} \omega$ (cyan),
(c) $\alpha_{xy}^{\rm ee} \omega Z_0$ (black) and $\alpha_{yx}^{\rm mm} \omega Z_0^{-1}$ (pink), 
(d) $\alpha_{yx}^{\rm em} \omega$ (green) and $\alpha_{xy}^{\rm me} \omega$ (purple).
(e) MO and (f) moving effects 
of the $y$-axis-oriented metamolecules (blue) 
are compared with those of the $x$-axis-oriented metamolecules (red). 
Solid and dashed lines present real and imaginary parts, 
respectively.
External magnetic field $\mu_{0} H_{\rm ext}$ is 400 mT.  
}
\label{fig:xy-ept}
\end{figure}

Figures \ref{fig:xy-ept}(a)-\ref{fig:xy-ept}(d) present 
extracted effective polarizability tensors 
of the $y$-axis-oriented metamolecule 
at 400 mT.
Solid and dotted lines 
correspond to the real and imaginary parts, respectively.
In Fig. \ref{fig:xy-ept}(a), 
$\alpha_{yy}^{\rm ee} \omega Z_0$  (red) 
corresponding to the diagonal part of electric susceptibility
shows electric dipole resonance at 9.3 GHz and 12.2 GHz. 
Contrastingly, 
$\alpha_{xx}^{\rm mm} \omega Z_0^{-1}$ (blue) 
corresponding to the diagonal part of magnetic susceptibility 
exhibits no magnetic dipole mode.
This is consistent with the fact that 
as AC electric fields in the $y$-direction, $E_y$, 
of TE$_{10}$ mode of microwaves in the waveguide 
excite electric dipole resonance $p_y$ in the Cu chiral meta-atom,
i.e., chiral resonance.
The chiral resonance $p_y$ 
induces large AC current in the Cu chiral meta-atom 
by the the Amp\`{e}re's circuital law,
resulting in magnetic dipole mode $m_y$ in the $y$-direction.
This corresponds to chiral-type bianisotropy, 
in which  $p_y$ is converted to $m_y$ (vice versa), 
by chirality. 
Indeed, 
as represented by $\alpha_{yy}^{\rm me} \omega$ (gold) 
and and $\alpha_{xx}^{\rm em} \omega$ (cyan) in Fig. \ref{fig:xy-ept}(b), 
the chiral-type bianisotropy is observed at 9.3 and 12.2 GHz.

The converted magnetic dipole mode $m_y$ 
is rotated around $\mu_{0} H_{\rm ext}$ along to the $z$-direction (MO effects) 
due to magnetization precession 
in the YIG magnetic meta-atom. 
Figure \ref{fig:xy-ept}(c) presents 
the off-diagonal part of electric susceptibility,
$\alpha_{xy}^{\rm ee}  \omega Z_0$ (black), 
and of magnetic susceptibility, 
$\alpha_{yx}^{\rm mm} \omega Z_0^{-1}$ (pink).
As in Fig. \ref{fig:xy-ept}(c), 
MO effects 
represented by magnetic $\alpha_{yx}^{\rm mm} \omega Z_0^{-1}$ 
are observed at 9.3 and 12.2 GHz.
In this way, 
$p_y$ in the $y$-direction
causes $m_x$ in the $x$-direction; 
here we obtain ME coupling $\alpha^{\rm me}$, 
which is defined by $m_x = \alpha_{xy}^{\rm me} E_y$ 
and assigned to the moving-type bianisotropy, 
i.e., the synthetic motion. 
Indeed in Fig. \ref{fig:xy-ept}(d), 
the large moving-type bianisotropy
represented by $\alpha_{yx}^{\rm em} \omega$ (green) 
and $\alpha_{xy}^{\rm me} \omega$ (purple) 
is observed.
The large synthetic motion of MChPs 
results in one-way transparency 
demonstrated in Figs. \ref{fig:s21s12} and \ref{fig:2dplot}(b).

Contrastingly in the $x$-axis-oriented metamolecule,
the $x$-direction AC magnetic fields of microwave photons, $H_x$,
give rise to electric dipole mode in the $x$-direction, $p_x$, 
by the Faraday's law of electromagnetic induction (chiral-type bianisotropy).
As the metamolecule has 
no intrinsic off-diagonal part of electric susceptibility, 
MO effects are not expected.
Nonetheless,  
$p_x$ possibly induces 
inhomogeneous magnetization in the YIG magnetic meta-atom. 
Therefore, 
the meta-atom smaller than the wavelength of microwaves 
is likely to have a small off-diagonal part of electric susceptibility.
Due to the small MO effects, 
$p_x$ is slightly rotated, 
resulting in electric dipole mode in the $y$-direction, $p_y$. 
However,
owing to the small MO effects, 
the moving-type bianisotropy 
by the $x$-axis-oriented metamolecule 
is small.

\textit{Origin of one-way transparency} --
We reveal that 
a large MO effect in the $y$-oriented metamolecule 
is the origin of the large synthetic motion for one-way transparency; 
this is highlighted in Figs. \ref{fig:xy-ept}(e) and \ref{fig:xy-ept}(f).
Figure \ref{fig:xy-ept}(e) shows 
$\omega Z_0^{-1} (\alpha_{yx}^{\rm mm} - \alpha_{xy}^{\rm mm}) /2$ (blue)
and $\omega Z_0 (\alpha_{yx}^{\rm ee} - \alpha_{xy}^{\rm ee}) /2$ (red)
corresponding to the MO effects 
in the $y$-axis-oriented 
and $x$-axis-oriented metamolecules, respectively. 
Similarly in Fig. \ref{fig:xy-ept}(f), 
$\omega (\alpha_{yx}^{\rm em} + \alpha_{xy}^{\rm me}) /2$
corresponding to the synthetic moving effects 
by the $y$-axis-oriented metamolecule (blue)
is compared with that by the $x$-axis-oriented metamolecule (red).
Solid and dashed lines present real and imaginary parts, 
respectively.

Figure \ref{fig:xy-ept}(e) illustrates that
the MO effect in the $y$-axis-oriented metamolecules 
(blue, $\omega Z_0^{-1} (\alpha_{yx}^{\rm mm} - \alpha_{xy}^{\rm mm})/2$) 
is much larger than 
that in the $x$-axis-oriented metamolecules 
(red, $\omega Z_0 (\alpha_{xy}^{\rm ee} - \alpha_{yx}^{\rm ee})/2$).
Given that the synthetic motion is a combination 
between MO effects and the chiral-type bianisotropy, 
large MO effects result in 
large synthetic motion of MChPs 
in the $y$-axis-oriented metamolecule (blue).
This mechanism of the enhancement in the synthetic motion 
is applicable to natural chiral molecules \cite{Rikken1997} 
and multiferroic materials \cite{Kezsmarki2014,Kezsmarki2015,Toyoda2015}.

We verify that  
the origin of the microwave one-way transparency 
observed in Fig. \ref{fig:s21s12}(a)
is the massive synthetic motion of MChPs 
in the $y$-axis-oriented MCh metamolecule. 
The large synthetic motion of MChPs 
are inherently independent of microwave polarizations,
resulting in an advantage in realization of synthetic gauge fields, 
for example, the Lorentz force for electromagnetic waves \cite{Sawada05} 
in the free space. 
Last but not least, 
Fig. \ref{fig:2dplot}(a) highlights that 
the synthetic motion
in the upper branch of MChPs modes
is significantly enhanced 
and one-way transparency is obtained 
at a higher $\mu_{0} H_{\rm ext}$, 
where the MChP mode is close to the MC-MChP mode. 
This may indicate that 
interaction between MChP and MC-MChP 
becomes much stronger 
via Mie resonance 
in the YIG magnetic meta-atom, 
leading to generation of a chimera quasiparticle.

\textit{Conclusion} --
We experimentally demonstrate 
microwave one-way transparency 
by the MCh metamolecule 
at room temperature.
The experimental results 
are reproduced via numerical simulations.
Using effective polarizability tensor analyses 
of the numerical results,  
we reveal that 
the one-way transparency is caused by 
the synthetic motion of ultrastrongly-coupled MChPs 
enhanced by the combination of the chiral-type bianisotropy and large MO effects.
The enhancement mechanism of the synthetic motion 
of MChPs in metamolecules at room temperature 
is applicable to natural chiral molecules and multiferroic materials, 
and signifies an advancement
toward hybrid quantum systems, 
synthetic gauge fields acting on light, 
and chimera quasiparticles using metamaterials.

\textit{Acknowledgement} --
We thank 
H. Kurosawa for helping 
with the numerical simulation.
This work is financially supported by 
JSPS KAKENHI (JP24H02232, 23K13621, 22K14591) and
JST-CREST (JPMJCR2102) and JST-FOREST (JPMJFR246R).


\begin{thebibliography}{}\label{sec:TeXbooks}
\bibitem{Mills1974} D. L. Mills and E. Burstein, Polaritons: the electromagnetic modes of media, Rep. Prog. Phys. 37, 817 (1974).
\bibitem{Kockum2019} A. F. Kockum, A. Miranowicz, S. De Liberato, S. Savasta, and F. Nori, Ultrastrong coupling between light and matter, Nature Reviews Physics 1, 19 (2019).
\bibitem{Diaz2019} P. Forn-D\`{i}az, L. Lamata, E. Rico, J. Kono, and E. Solano, Ultrastrong coupling regimes of light-matter interaction, Reviews of Modern Physics 91, 025005 (2019).
\bibitem{Qin2024} W. Qin, A. F. Kockum, C. S. Mu\~{n}oz, A. Miranowicz, and F. Nori, Quantum amplification and simulation of strong and ultrastrong coupling of light and matter, Phys. Rep. 1078, 1 (2024).
\bibitem{Soykal2010} \"{O}. O. Soykal and M. E. Flatt\'{e}, Strong Field Interactions between a Nanomagnet and a Photonic Cavity, Phys. Rev. Lett. 104, 077202 (2010).
\bibitem{Huebl2013} H. Huebl, C. W. Zollitsch, J. Lotze, F. Hocke, M. Greifenstein, A. Marx, R. Gross, and S. T. B. Goennenwein, High Cooperativity in Coupled Microwave Resonator Ferrimagnetic Insulator Hybrids, Phys. Rev. Lett. 111, 127003 (2013).
\bibitem{Tabuchi2014} Y. Tabuchi, S. Ishino, T. Ishikawa, R. Yamazaki, K. Usami, and Y. Nakamura, Hybridizing Ferromagnetic Magnons and Microwave Photons in the Quantum Limit, Phys. Rev. Lett. 113, 083603 (2014).
\bibitem{Li2020} Y. Li, W. Zhang, V. Tyberkevych, W.-K. Kwok, A. Hoffmann, and V. Novosad, Hybrid magnonics: Physics, circuits, and applications for coherent information processing, Journal of Applied Physics 128, 130902 (2020).
\bibitem{Harder2021} M. Harder, B. M. Yao, Y. S. Gui, and C.-M. Hu, Coherent and dissipative cavity magnonics, Journal of Applied Physics 129, 201101 (2021).
\bibitem{Lachance-Quirion2019} D. Lachance-Quirion, Y. Tabuchi, A. Gloppe, K. Usami, and Y. Nakamura, Hybrid quantum systems based on magnonics, Applied Physics Express 12, 070101 (2019).
\bibitem{Golovchanskiy2021} I. A. Golovchanskiy, N. N. Abramov, V. S. Stolyarov, M. Weides, V. V. Ryazanov, A. A. Golubov, A. V. Ustinov, and M. Y. Kupriyanov, Ultrastrong photon-to-magnon coupling in multilayered heterostructures involving superconducting coherence via ferromagnetic layers, Sci. Adv. 7, eabe8638 (2021).
\bibitem{Walls_book} D. F. Walls and G. J. Milburn, {\it Quantum Optics} (Springer, Berlin 2008).  
\bibitem{Wang2019} Y.-P. Wang, J. W. Rao, Y. Yang, P.-C. Xu, Y. S. Gui, B. M. Yao, J. Q. You, and C.-M. Hu, Nonreciprocity and Unidirectional Invisibility in Cavity Magnonics, Phys. Rev. Lett. 123, 127202 (2019).
\bibitem{Zhang2020} X. Zhang, A. Galda, X. Han, D. Jin, and V. M. Vinokur, Broadband Nonreciprocity Enabled by Strong Coupling of Magnons and Microwave Photons, Phys. Rev. Appl. 13, 044039 (2020).
\bibitem{Tomita2014} S. Tomita, K. Sawada, A. Porokhnyuk, and T. Ueda, Direct Observation of Magnetochiral Effects through a Single Metamolecule in Microwave Regions, Physical Review Letters 113, 235501 (2014).
\bibitem{Tomita2017} S. Tomita, H. Kurosawa, K. Sawada, and T. Ueda, Enhanced magnetochiral effects at microwave frequencies by a single metamolecule, Physical Reviews B 95, 085402 (2017).
\bibitem{Tomita2018} S. Tomita, H. Kurosawa, T. Ueda, and K. Sawada, Metamaterials with magnetism and chirality, Journal of Physics D: Applied Physics 51, 083001 (2018).
\bibitem{Kurosawa2022} H. Kurosawa, S. Tomita, K. Sawada, T. Nakanishi, and T. Ueda, Unity-order magnetochiral effects exhibited by a single metamolecule, Opt. Expr. 30, 37066 (2022).
\bibitem{mita2025} K. Mita, T. Chiba, T. Kodama, T. Ueda, T. Nakanishi, K. Sawada, and S. Tomita, Ultrastrongly coupled and directionally nonreciprocal magnon polaritons in magnetochiral metamolecules, Phys. Rev. Appl. 23, L011004 (2025).
\bibitem{Huidobro2019} P. A. Huidobro, E. Galiffi, S. Guenneau, and J. B. Pendry, Fresnel drag in space-time-modulated metamaterials, Proc. Natl. Acad. Sci. U.S.A. 116, 24943 (2019).
\bibitem{asadchy2020} V. S. Asadchy, M. S. Mirmoosa, A. D\'{i}az-Rubio, S. Fan, and S. A. Tretyakov, Tutorial on Electromagnetic Nonreciprocity and its Origins, Proceedings of the IEEE 108, 1684 (2020).
\bibitem{Barron1984} L. D. Barron and J. Vrbancich, Magneto-chiral birefringence and dichroism, Mol. Phys. 51, 715 (1984).
\bibitem{Rikken1997} G. L. J. A. Rikken and E. Raupach, Observation of magneto-chiral dichroism, Nature 390, 493 (1997).
\bibitem{Kezsmarki2014} I. K\'{e}zsm\'{a}rki, D. Szaller, S. Bord\'{a}cs, V. Kocsis, Y. Tokunaga, Y. Taguchi, H. Murakawa , Y. Tokura, H. Engelkamp, T. R\~{o}\~{o}m, and U. Nagel, One-way transparency of four-coloured spin-wave excitations in multiferroic materials, Nat. Commun. 5, 3203 (2014).
\bibitem{Kezsmarki2015} I. K\'{e}zsm\'{a}rki, U. Nagel, S. Bord\'{a}cs, R. S. Fishman, J. H. Lee, H. T. Yi, S.-W. Cheong, and T. R\~{o}\~{o}m, Optical Diode Effect at Spin-Wave Excitations of the Room-Temperature Multiferroic BiFeO$_3$, Phys. Rev. Lett. 115, 127203 (2015).
\bibitem{Toyoda2015} S. Toyoda, N. Abe, S. Kimura, Y. H. Matsuda, T. Nomura, A. Ikeda, S. Takeyama, and T. Arima, One-Way Transparency of Light in Multiferroic CuB$_2$O$_4$, Phys. Rev. Lett. 115, 267207 (2015).
\bibitem{Kuzmenko2015} A. M. Kuzmenko, V. Dziom, A. Shuvaev, An. Pimenov, M. Schiebl, A. A. Mukhin, V. Yu. Ivanov, I. A. Gudim, L. N. Bezmaternykh, and A. Pimenov, Large directional optical anisotropy in multiferroic ferroborate, Phys. Rev. B 92, 184409 (2015).
\bibitem{nagaosa2024} N. Nagaosa and Y. Yanase, Nonreciprocal Transport and Optical Phenomena in Quantum Materials, Annu. Rev. Condens. Matter Phys. 15, 63 (2024).
\bibitem{Sawada05} K. Sawada and N. Nagaosa, Optical Magnetoelectric Effect in Multiferroic Materials: Evidence for a Lorentz Force Acting on a Ray of Light, Phys. Rev. Lett. 95, 237402 (2005).
\bibitem{Mirmoosa2014} M. S. Mirmoosa, Y. Ra'di, V. S. Asadchy, C. R. Simovski, and S. A. Tretyakov, Polarizabilities of Nonreciprocal Bianisotropic Particles, Phys. Rev. Appl. 1, 034005 (2014).
\bibitem{Alaee2015} R. Alaee, M. Albooyeh, M. Yazdi, N. Komjani, C. Simovski, F. Lederer, and C. Rockstuhl, Magnetoelectric coupling in nonidentical plasmonic nanoparticles: Theory and applications, Phys. Rev. B 91, 115119 (2015).
\bibitem{Yazdi2016} M. Yazdi and N. Komjani, Polarizability calculation of arbitrary individual scatterers, scatterers in arrays, and substrated scatterers, J. Opt. Soc. Am. B 33, 491 (2016).
\bibitem{Kodama2024} T. Kodama, T. Nakanishi, K. Sawada, and S. Tomita, Pure moving optical media consisting of magnetochiral metasurfaces, Opt. Mater. Expr. 14, 2499 (2024).
\bibitem{SM} See Supplemental Material [url] for numerical simulation setup using COMSOL Multiphysics; Mie resonance by YIG magnetic meta-atom; effective polarizability tensor of bianisotropic media; and effective polarizability tensor analyses. The Supplemental Material also contains Refs. \cite{asadchy2018,simovski_book,Kong,Hecht,caloz2020,tellegen1948}.
\bibitem{asadchy2018} V. S. Asadchy, A. D\'{i}az-Rubio, and S. A. Tretyakov, Bianisotropic metasurfaces: physics and applications, Nanophotonics 7, 1069 (2018). 
\bibitem{simovski_book} C. Simovski and S. Tretyakov, {\it An introduction to Metamaterials and Nanophotonics} (Cambridge University Press, Cambridge, 2020).
\bibitem{Kong} J. A. Kong, {\it Electromagnetic Wave Theory} (EMW Publishing, Cambridge, 2005).
\bibitem{Hecht} E. Hecht, {\it Optics} (Pearson Education Limited, Harlow, 2013).
\bibitem{caloz2020} C. Caloz and A. Sihvola, Electromagnetic chirality, part 1 : Microscopic perspective, IEEE Antennas Propag. 62, 58 (2020).
\bibitem{tellegen1948} B. D. Tellegen, The gyrator, a new electric network element, Philips Res. Rep. 3, 81-101 (1948). 
\bibitem{BH} C. F. Bohren and D. R. Huffman, {\it Absorption and Scattering of Light by Small Particles} (Wiley-Interscience, New York, 1983).

\end{thebibliography}

\end{document}